\documentclass[10pt,journal]{IEEEtran}

\usepackage[pdftex]{graphicx}
\usepackage{algorithm}
\usepackage{color}
\usepackage{caption}
\usepackage{algpseudocode}

\definecolor{ForestGreen}{RGB}{162,52,0}

\usepackage{xcolor}
\usepackage{amsmath}
\usepackage{amssymb}
\usepackage{latexsym}
\usepackage{CJK}
\usepackage{url}
\usepackage{array}
\usepackage{ragged2e}
\usepackage{makecell}
\usepackage{multirow}
\usepackage{booktabs}
\usepackage{threeparttable}

\ifCLASSOPTIONcompsoc
  \usepackage[nocompress]{cite}
\else
  \usepackage{cite}
\fi

\ifCLASSINFOpdf
\else
\fi

\ifCLASSOPTIONcompsoc
  \usepackage[caption=false,font=footnotesize,labelfont=sf,textfont=sf]{subfig}
\else
  \usepackage[caption=false,font=footnotesize]{subfig}
\fi

\begin{document}

\title{Space-Air-Ground Integrated Multi-domain Network Resource Orchestration based on Virtual Network Architecture: a DRL Method}

\author{Peiying Zhang, Chao Wang, Neeraj Kumar,~\IEEEmembership{Senior Member,~IEEE}, and Lei Liu

\thanks{This work was supported in part by the National Natural Science Foundation of China under Grant 62001357; in part by the Major Scientific and Technological Projects of CNPC under Grant ZD2019-183-006; in part by the Guangdong Basic and Applied Basic Research Foundation under Grant 2020A1515110496 and Grant 2020A1515110079; and in part by the Shandong Provincial Natural Science Foundation, China, under Grant ZR2020MF006; and in part by the Open Foundation of State key Laboratory of Networking and Switching Technology (Beijing University of Posts and Telecommunications) under Grant SKLNST-2021-1-17. \textit{(Corresponding authors: Neeraj Kumar and Peiying Zhang)}.}
\thanks{Peiying Zhang is with the College of Computer Science and Technology, China University of Petroleum (East China), Qingdao 266580, China, and also with the State Key Laboratory of Networking and Switching Technology, Beijing University of Posts and Telecommunications, Beijing 100876, China. E-mail: zhangpeiying@upc.edu.cn.}
\thanks{Chao Wang is with the College of Computer Science and Technology, China University of Petroleum (East China), Qingdao 266580, China. E-mail: wangch\_upc@qq.com.}
\thanks{Neeraj Kumar is with the Department of Computer Science and Engineering, Thapar Institute of Engineering and Technology, Deemed to be University, Patiala 147004, India, also with the Department of Computer Science and Information Engineering, Asia University, Taichung 41354, Taiwan, and also with the School of Computer Science, University of Petroleum and Energy Studies, Dehradun 248007, India. Email: neeraj.kumar@thapar.edu.}
\thanks{Lei Liu is with the State Key Laboratory of Integrated Services Networks, Xidian University, Xi'an 710071, China. Email: leiliu@xidian.edu.cn.}
}

\markboth{IEEE Transactions on Intelligent Transportation Systems, ~Vol.~XX, No.~XX, XX~2021}
{Shell \MakeLowercase{\textit{Zhang et al.}}: Bare Demo of IEEEtran.cls for Computer Society Journals}

\maketitle
\begin{abstract}
Traditional ground wireless communication networks cannot provide high-quality services for artificial intelligence (AI) applications such as intelligent transportation systems (ITS) due to deployment, coverage and capacity issues. The space-air-ground integrated network (SAGIN) has become a research focus in the industry. Compared with traditional wireless communication networks, SAGIN is more flexible and reliable, and it has wider coverage and higher quality of seamless connection. However, due to its inherent heterogeneity, time-varying and self-organizing characteristics, the deployment and use of SAGIN still faces huge challenges, among which the orchestration of heterogeneous resources is a key issue. Based on virtual network architecture and deep reinforcement learning (DRL), we model SAGIN's heterogeneous resource orchestration as a multi-domain virtual network embedding (VNE) problem, and propose a SAGIN cross-domain VNE algorithm. We model the different network segments of SAGIN, and set the network attributes according to the actual situation of SAGIN and user needs. In DRL, the agent is acted by a five-layer policy network. We build a feature matrix based on network attributes extracted from SAGIN and use it as the agent training environment. Through training, the probability of each underlying node being embedded can be derived. In test phase, we complete the embedding process of virtual nodes and links in turn based on this probability. Finally, we verify the effectiveness of the algorithm from both training and testing.
\end{abstract}


\IEEEpeerreviewmaketitle

\section{Introduction}\label{part1}

In recent years, with the vigorous development of the artificial intelligence (AI) industry, intelligent transportation systems (ITS) have entered a stage of rapid development \cite{n5,r1}. Among them, as the main part of ITS, vehicular communication networks (VCNs) mainly rely on the communication services provided by 802.11p networks and cellular networks, which can complete vehicular functions such as road safety, entertainment interaction and location awareness to a certain extent \cite{e2,r2}. The development potential of VCN is huge. It is estimated that the number of vehicles connected to the Internet will reach 286 million by 2025. However, the deployment of VCN is facing a series of inevitable problems. First of all, 802.11p networks and cellular networks only provide dedicated short-distance communication services. The rapid movement of vehicles may cause frequent terminals on the network, thereby reducing service quality \cite{l2}. Secondly, the deployment of ground communication facilities (base stations (BSs), roadside units (RSUs)) is expensive and takes a long time to deploy \cite{r3,z4}. It is impossible to achieve high coverage in rural or remote mountainous areas. Finally, ground communication facilities are easily damaged by natural disasters such as earthquakes or floods, and cannot provide stable communication services for vehicles at any time \cite{l1}. Therefore, VCN deployment, coverage and capacity issues still need to be resolved urgently \cite{n2,n6}. Radio network resource management faces severe challenges, including storage, spectrum, computing resource allocation, and joint allocation of multiple resources \cite{jcx1,jcx2}. With the rapid development of communication networks, the integrated space-ground network has also become a key research object \cite{jcx3}.

\begin{figure}[!h]
\centering
\includegraphics[width=1\columnwidth]{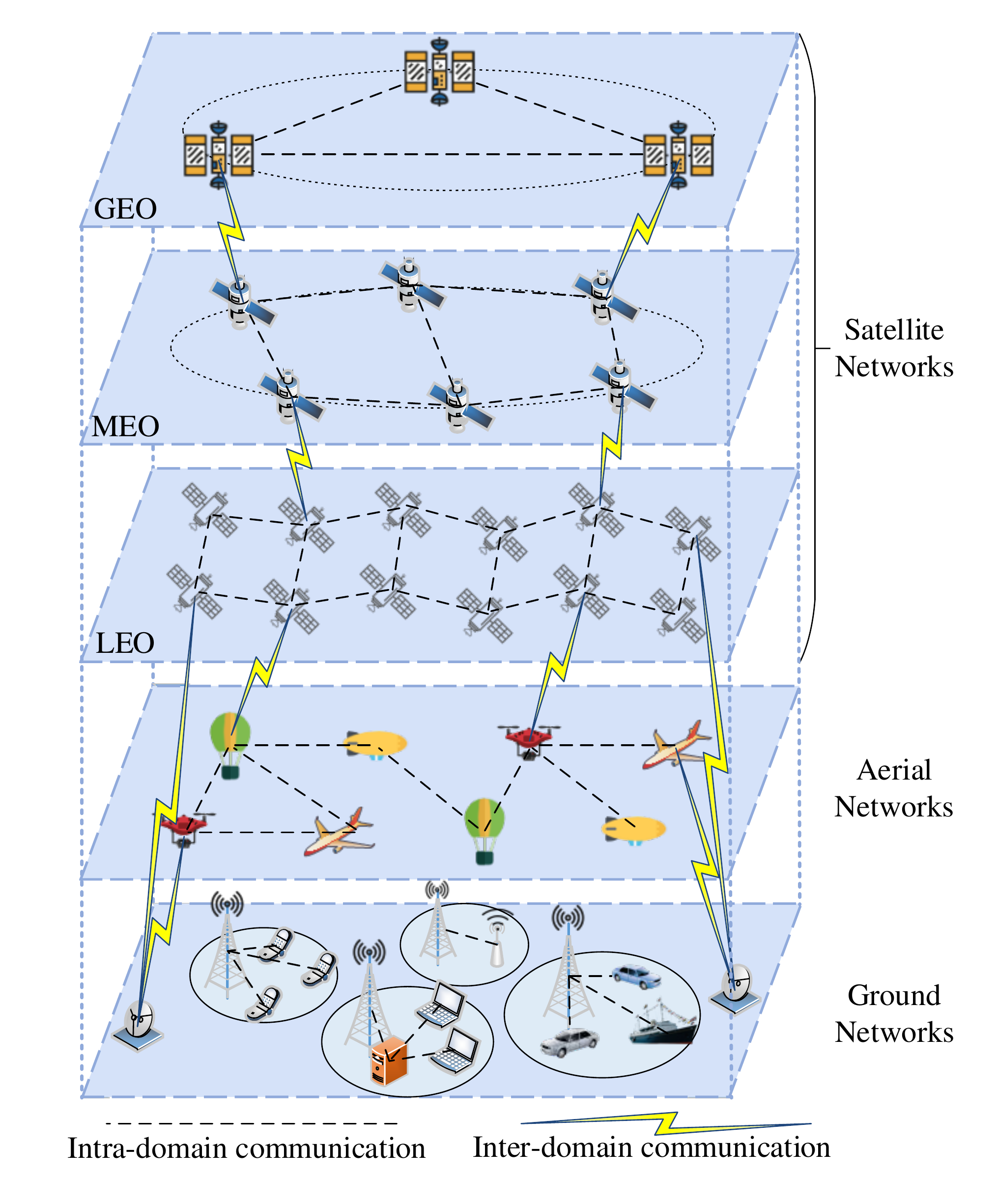}
\caption{A typical space-air-ground integrated network architecture.}
\label{fig_1}
\end{figure}

Space-air-ground integrated networks (SAGIN) can provide three-dimensional network connection for vehicles anytime and anywhere, which has become the key research direction of the next generation of ITS \cite{b2}. For example, Tesla plans to launch a certain number of commercial satellites to provide global connectivity services for new energy vehicles. The mTenna equipped with Toyota's Mirai car can provide it with a transfer rate of $50MB/s$. Google and Facebook also plan to deploy balloons and unmanned aerial vehicles (UAVs) to provide Internet services in remote areas, respectively. As a promising network architecture, SAGIN can provide seamless global connectivity and efficient and reliable low latency services for emerging applications including VCN \cite{e1}. SAGIN is essentially a layered network architecture, which is mainly composed of three network segments, as shown in Fig. \ref{fig_1}. Satellites can be divided into three categories according to the height above the ground: geosynchronous orbit (GEO), medium earth orbit (MEO) and low earth orbit (LEO) satellites. Air networks can be divided into high altitude platform (HAP) and low altitude platform (LAP). UAVs, balloons and airships are its main components \cite{e3}. Ground networks mainly refer to traditional communication networks such as cellular networks and wireless local area networks (WLAN).

SAGIN is a layered network architecture, and different network segments are quite different. The network nodes composed of satellites or UAVs are always in a mobile state, so SAGIN has inherent characteristics such as heterogeneity, time-varying and self-organizing \cite{e4}. SAGIN is restricted by many factors such as traffic distribution, routing scheduling, power control, spectrum allocation, and load balancing \cite{e6}. Among them, the allocation and scheduling of heterogeneous physical network resources is a key issue. A reliable idea is to adopt a new architecture to enhance SAGIN, focusing on solving the allocation and scheduling of SAGIN's heterogeneous physical network resources, i.e., the problem of network resource orchestration. Network virtualization (NV) is a technology that logically abstracts physical networks \cite{n3,n1}. It can solve the problem of resource allocation in heterogeneous networks by providing intelligent and flexible management and orchestration systems. Virtual network embedding (VNE) is the core issue of NV research, and its essence is the orchestration of network resources \cite{z1}. Therefore, the resource allocation problem of SAGIN can be transferred to the research of VNE algorithm. In SAGIN, physical network resources may come from different network segments, so we consider implementing a multi-domain VNE algorithm based on a virtual network architecture.

AI technology solves many problems in daily production and life with its superior performance, especially in the field of perception and decision-making high-dimensional space problems \cite{z5}. The rapid development and universal application of deep learning (DL) and reinforcement learning (RL) are the key to the success of AI \cite{n4,z3}. The former has strong perception ability, while the latter has strong decision-making ability. The product of the combination of the two is deep reinforcement learning (DRL). DRL is essentially an end-to-end perception and control system, which has strong versatility and is usually used to solve decision-making problems in high-dimensional space. Scholars have already used this technology to improve network performance \cite{a1,a2}. VNE is NP-hard \cite{z2}. DRL has better performance than optimization methods or heuristic methods when solving such problems \cite{b1,e5}. Therefore, we consider using DRL methods to optimize the resource scheduling problem of SAGIN.

This paper has done the following main work.
\begin{enumerate}
\item In order to improve the efficiency and rationality of the allocation of heterogeneous network resources in SAGIN, based on the virtual network architecture, we model the resource scheduling problem of SAGIN as a multi-domain VNE problem, and provide a solution for resource allocation of SAGIN from the perspective of VNE.
\item We use DRL to improve the performance of the multi-domain VNE algorithm. Specifically, we use a self-built policy network as an agent, and form a feature matrix by extracting SAGIN resource attributes to provide an environment for agent training. DRL method can deduce the node embedding probability, and then complete the entire multi-domain VNE algorithm.
\item We verify the performance of the proposed algorithm through experimental simulations. According to the actual network characteristics of different network segments of SAGIN, we set differentiated network attributes for the network topology. Experimental results show that the proposed algorithm performs well in multiple network performance.
\end{enumerate}

The main content of the rest of this paper is as follows. Section \ref{part2} reviews the related research work carried out on SAGIN, including SAGIN based on virtual network architecture. Section \ref{part3} describes related issues and system models. Section \ref{part4} gives the constraints and performance indicators of the algorithm. Section \ref{part5} introduces the algorithm realization process. We showed and analyzed the experimental results in Section \ref{part6}. Section \ref{part7} summarizes the full paper.

\section{Related Work}\label{part2}

\subsection{Overview of Research Status of SAGIN Technology}

SAGIN, as an important form of future wireless network communication system, has become a research hotspot in industry and academia. Scholars have carried out a lot of research on SAGIN related technologies. Reference \cite{a3} and \cite{a4} have carried out research from satellite system and UAV communication network respectively. They focused on summarizing a series of problems (dynamic topology, energy loss and capacity limitation) faced by satellite communications and UAV communications. Spectrum allocation, traffic offloading, routing strategy and system integration are key issues in SAGIN research. Li et al. \cite{a5} developed a spectrum allocation scheme for cognitive satellite networks. This solution used Bayesian equalization as the final spectrum allocation strategy, and improved spectrum efficiency by making full use of spectrum resources and overall user demand. The authors of \cite{a6} studied the spectrum allocation problem when multiple UAVs were used as relay nodes. The authors aimed to maximize the service efficiency of the integrated IoT network system, and solved the joint optimization problem of bandwidth allocation, gateway selection, and UAV deployment based on simulated annealing and continuous convex planning. In order to provide an economical, reliable, and efficient resource management solution for air vehicles, Varasteh et al. \cite{a7} modeled routing and service placement problems as virtual machine placement problems. After weighing different optimization solutions, the authors decided on the routing, service placement and service migration of the aircraft in SAGIN, realizing the dynamic adjustment of the service network. Ruan et al. \cite{a8} studied the issue of spectrum efficiency between satellite networks and ground networks. The authors proposed an adaptive transmission scheme with symbol error rate (SER) constraints. Finally, they took the SER as a constraint and discussed the trade-off between energy efficiency and spectrum efficiency.

\subsection{Research Status of SAGIN based on Virtual Network Architecture}

Virtual network architecture has begun to try to apply in SAGIN. As a future network architecture, virtual network has significant advantages in development and management of heterogeneous resources, and have been favored by researchers. As excellent representatives of NV, software defined network (SDN) and network function virtualization (NFV) are considered to be enabling technologies for the flexible and effective integration of heterogeneous networks, and can provide innovative solutions for the orchestration of heterogeneous network resources.

Authors of \cite{a9} proposed a software defined SAGIN architecture based on reviewing the motivation and challenges of SAGIN integration. In order to protect the traditional services in different segmented networks, the authors used network slicing to slice the resources of each network segment, and then put the available resources into a public resource pool, which provided reliable services for the Internet of Vehicles. In order to optimize the load balancing of network communication, reference \cite{a10} studied a software defined SAGIN routing algorithm. Based on the characteristics of the SDN model and the dynamic changes of SAGIN topology, the authors considered the multidimensionality of resources and energy consumption, which effectively reduced the end-to-end delay and packet loss rate. Wang et al. \cite{a11} proposed a SAGIN reconfigurable service framework based on service function chain (SFC). The framework modeled the realization of SFC and virtual network functions (VNFs) as integer nonlinear programming problems. The authors proposed a heuristic greedy algorithm to balance the resource consumption of different network nodes. The results proved that this algorithm can improve resource utilization efficiency. Du et al. \cite{a12} studied spectrum sharing and interference control technology based on SDN. They proposed a spectrum sharing and service offloading mechanism to realize the cooperative relationship between the ground BSs and the beam group of the satellite ground communication system. In this mode, the communication between satellites and the ground effectively realized frequency sharing and traffic offloading.

\subsection{Research Status Analysis}

Based on the comprehensive analysis of SAGIN related research, it is found that they all have the following problems.
\begin{enumerate}
\item In the relevant technical research field of SAGIN, they only study space, air, and ground one of the network segments or any two network segments combined with related technologies, and do not pay attention to the integration of space-air-ground three-dimensional network segments.
\item In the research field of SAGIN technology based on virtual network architecture, the existing work usually adopt optimization methods or heuristic methods to solve SAGIN resource management problems. They do not apply AI algorithms to SAGIN resource orchestration problems.
\item The existing work only analyze the possible impact of the heterogeneity and time-varying of SAGIN on the network resource allocation, and do not model the physical network resources of different network segments, so it cannot fully reflect the dynamic changes of network resources when SAGIN provides services for end users.
\end{enumerate}
Therefore, on the basis of analyzing the inherent characteristics of SAGIN, such as heterogeneity, time-varying and self-organizing, this paper proposes a SAGIN resource orchestration algorithm based on virtual network architecture and DRL method, which is essentially a multi-domain VNE algorithm.

\section{Problem Description and System Model}\label{part3}

\subsection{Problem Description}

One of the most prominent features of SAGIN is time-varying. Since satellites and aerial vehicles are constantly moving, the network topology will always change. For example, when a vehicle is receiving the positioning service provided by satellite $A$, but due to the revolution of the satellite, the vehicle leaves the service coverage area of satellite $A$ and enters the service coverage area of satellite $B$ instead. Then the mode of providing network resources has changed. In addition, due to the heterogeneity of SAGIN, the network resources of different network segments are also heterogeneous. Satellite nodes or aerial vehicle nodes have small capacity due to volume limitations, so their computing resources are often limited. It should be noted that the delay attributes of channel links in different network segments are often quite different \cite{r4,r5}.

We assume that the network topology of different network segments of SAGIN is relatively unchanged within a certain period of time, and end users are always within the service range of one or several satellites or aircrafts. Under the virtual network architecture, the network resource request sent by the end user to SAGIN forms a virtual network request (VNR). SAGIN will allocate network resources of different network segments according to the actual needs of the VNR. We regard $T$ as the reconstruction period of SAGIN resources, and the VNR may arrive at any point within $T$. The process of a VNR request service will not be interrupted by the deadline $T$. If the VNR cannot be completed within a period of time, it will continue to complete the request in the next reconstruction period $T$. When the VNR leaves, the network resources occupied by it will be released. The ultimate goal is to increase the revenue of network service providers on the basis of accepting as many VNRs as possible. Therefore, the problem of VNE across multiple network domains is finally formed. TABLE \ref{tab_1} summarizes the notations used in the multi-domain VNE problem of SAGIN based on the virtual network architecture.

\begin{table}
\centering
\caption{Notations}
\renewcommand\arraystretch{1.5}
\begin{tabular}{|p{6mm}|p{9mm}|p{60mm}|}
\hline
\multicolumn{2}{|l|}{Notations} & Descriptions  \\
\hline
\multicolumn{2}{|c|}{$G^P$} & Physical networks \\
\hline
\multirow{3}{*}{$N^P$} & $N_S^P$ & Satellite network nodes \\
\cline{2-3}
 & $N_A^P$ & Air network nodes \\
\cline{2-3}
 & $N_G^P$ & Ground network nodes \\
\hline
\multirow{6}{*}{$E^P$} & $E_S^P$ & Satellite network links \\
\cline{2-3}
 & $E_A^P$ & Air network links \\
\cline{2-3}
 & $E_G^P$ & Ground network links \\
\cline{2-3}
 & $E_{S,A}^P$ & Satellite network and air network inter-domain links \\
\cline{2-3}
 & $E_{S,G}^P$ & Satellite network and ground network inter-domain links \\
\cline{2-3}
 & $E_{A,G}^P$ & Air network and air network inter-domain links \\
\hline
\multirow{9}{*}{$A^P$} & $CPU_{N_S^P}$ & Satellite network nodes computing resources \\
\cline{2-3}
 & $CPU_{N_A^P}$ & Air network nodes computing resources \\
\cline{2-3}
 & $CPU_{N_G^P}$ & Ground network nodes computing resources \\
\cline{2-3}
 & $BW_{E_S^P}$ & Satellite network links bandwidth resources \\
\cline{2-3}
 & $BW_{E_A^P}$ & Air network links bandwidth resources \\
\cline{2-3}
 & $BW_{E_G^P}$ & Ground network links bandwidth resources \\
\cline{2-3}
 & $D_{E_S^P}$ & Space network links delay attributes \\
\cline{2-3}
 & $D_{E_A^P}$ & Air network links delay attributes \\
\cline{2-3}
 & $D_{E_G^P}$ & Ground network links delay attributes \\
\hline
\multicolumn{2}{|c|}{$G^V$} & Virtual network requests \\
\hline
\multicolumn{2}{|c|}{$N^V$} & Virtual nodes \\
\hline
\multicolumn{2}{|c|}{$E^V$} & Virtual links \\
\hline
\multirow{3}{*}{$A^V$} & $CPU_{N^V}$ & Computing resource requests of virtual nodes \\
\cline{2-3}
 & $BW_{E^V}$ & Bandwidth resource requests of virtual links \\
\cline{2-3}
 & $D_{E^V}$ & Delay requests of virtual links \\
\hline
\end{tabular}
\label{tab_1}
\end{table}

\subsection{System Model}

\subsubsection{Physical Network Model}

The physical network is modeled as an undirected weighted graph $G^P=\{N^P, E^P, A^P\}$, where $N^P$ represents the network node set of SAGIN, $E^P$ represents the network link set of SAGIN, and $A^P$ represents the network attribute set of SAGIN. Network node set $N^P=\{N_S^P, N_A^P, N_G^P\}$, where $N_S^P$ is the satellite node set, $N_A^P$ is the aerial node set, and $N_G^P$ is the ground node set. Network link set $E=\{E_S^P, E_A^P, E_G^P, E_{S,A}^P, E_{S,G}^P, E_{A,G}^P\}$, where $E_S^P$, $E_A^P$ and $E_G^P$ are the physical links among satellite nodes, air nodes and ground nodes respectively. In particular, $E_{S,A}^P$ is the set of inter domain links between satellite nodes and air nodes, $E_{S,G}^P$ is the set of inter domain links between satellite nodes and ground nodes, and $E_{A,G}^P$ is the set of links between aerial nodes and ground nodes. Network attribute set $A^P=\{CPU_{N_S^P},CPU_{N_A^P},CPU_{N_G^P},BW_{E_S^P},BW_{E_A^P},BW_{E_G^P},D_{E_S^P},\\D_{E_A^P},D_{E_G^P}\}$, where $CPU_{N_S^P}$ represents the computing resource attributes of satellite nodes, $CPU_{N_A^P}$ represents the computing resource attributes of air nodes, and $CPU_{N_G^P}$ represents the computing resource attributes of ground nodes. $BW_{E_S^P}$, $BW_{E_A^P}$ and $BW_{E_G^P}$ are the bandwidth resource attribute of satellite network links, air network links and ground network links respectively. $D_{E_S^P}$, $D_{E_A^P}$ and $D_{E_G^P}$ are the delay attribute of satellite network links, air network and ground network links respectively. We use $\{(N_m^P,N_n^P) \in E^P|N_m^P,N_n^P \in N^P\}$ to indicate that there are links between nodes $N_m^P$ and $N_n^P$. Thus, the bandwidth resource attribute between nodes $N_m^P$ and $N_n^P$ can be expressed as $BW_{N_m^P,N_n^P}^P$, and the single hop delay between nodes $N_m^P$ and $N_n^P$ can be expressed as $D_{N_m^P,N_n^P}^P$.

\subsubsection{Virtual Network Model}

VNRs are modeled as undirected weighted graph $G^V=\{N^V,E^V,A^V\}$, where $N^V$ represents the virtual node set, $E^V$ represents the virtual link set, and $A^V$ represents the attribute set of VNRs. Attribute set $A^V=\{CPU_{N^V},BW_{E^V},D_{E^V}\}$, where $CPU_{N^V}$ represents the computing resource requirements of virtual nodes, $BW_{E^V}$ represents the bandwidth resource requirements of virtual links, and $D_{E^V}$ represents the delay requirements of virtual links. In particular, we use $\{(N_j^V,N_k^V) \in E^V|N_j^V,N_k^V \in N^V\}$ to indicate that there is a link between virtual nodes $N_j^V$ and $N_k^V$. Thus, the bandwidth resource requirement between virtual nodes $j$ and $k$ can be expressed as $BW_{N_j^V,N_k^V}^V$, and the delay requirement between virtual nodes $j$ and $k$ can be expressed as $D_{N_j^V,N_k^V}^V$.

\begin{figure}[!h]
\centering
\includegraphics[width=1\columnwidth]{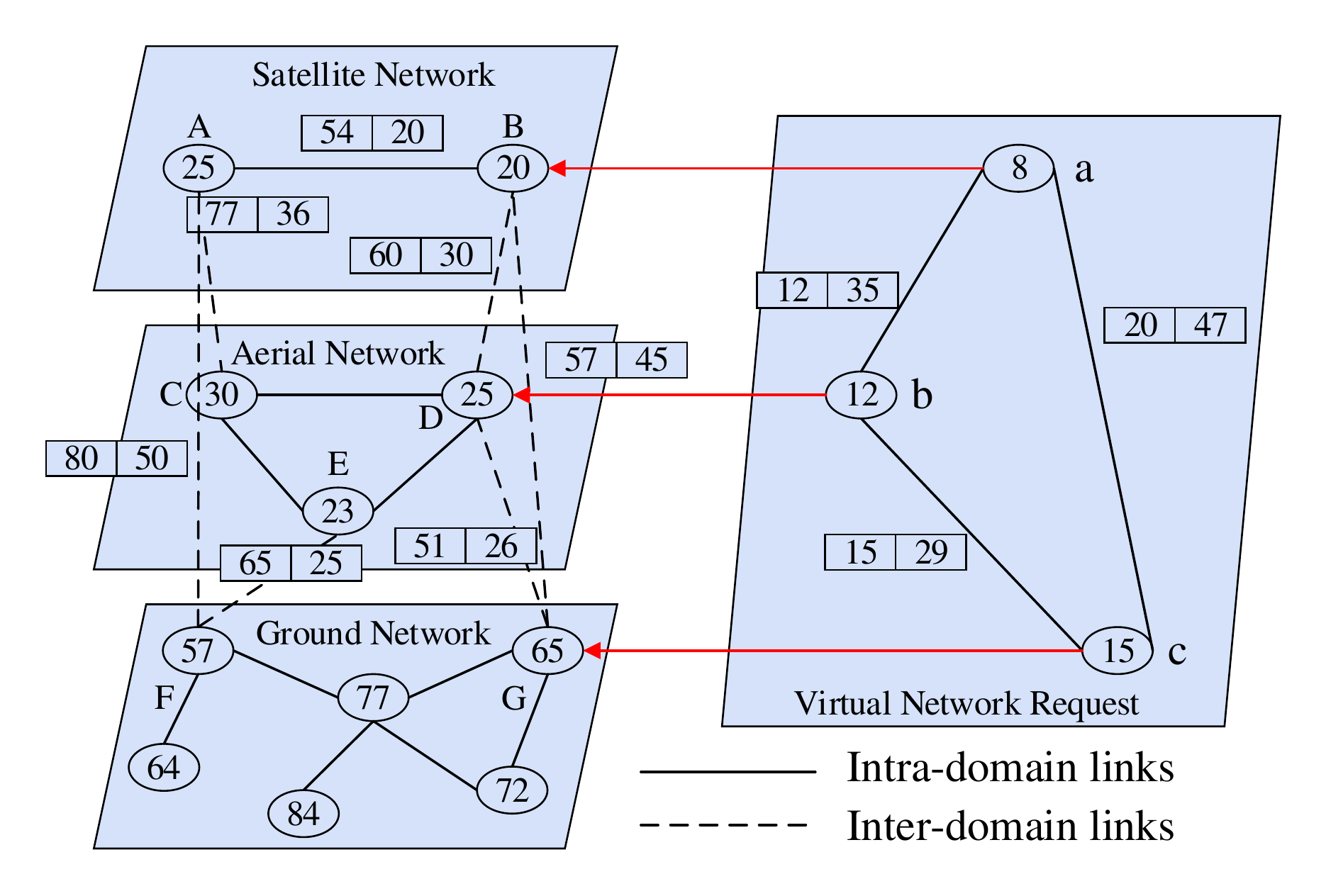}
\caption{Example of cross-domain VNE. On the left is a layered SAGIN architecture, and on the right is a VNR.}
\label{fig_2}
\end{figure}

Fig. \ref{fig_2} shows an example of a VNR embedded in SAGIN. The ellipses in the figure represent network nodes, and the connections between nodes represent network links. In SAGIN, the number on the node represents the amount of computing resources, and the number on the link represents the amount of bandwidth resources and the delay value respectively. In VNR, the number on the node represents the computing resource demand, and the number on the link represents the bandwidth resource demand and the maximum delay value respectively.

In the feasible VNE scheme, virtual node $a$ is mapped to satellite node $B$, virtual node $b$ is mapped to aerial node $D$, and virtual node $c$ is mapped to ground node $G$. The CPU resource capacity of the mapped physical node meets the requirements of the virtual node. At the same time, the link resource conditions between each node are also satisfied. If virtual node $a$ is mapped to satellite node $A$, virtual node $b$ is mapped to aerial node $C$, and virtual node $c$ is mapped to ground node $F$. At this time, the delay of the inter domain link between $A$ and $C$ is greater than the delay requirement of the virtual link between $a$ and $b$, and the delay of the inter domain link between $A$ and $F$ is also greater than the delay requirement of the virtual link between $a$ and $c$, so this is not a feasible VNE scheme.

\section{Constraints and Performance Indicators}\label{part4}

\subsection{Attribute Constraints}

We use the binary variable $x_{n^v,n^p}$ to indicate whether the virtual node $n^v$ is embedded on the physical node $n^p$, as shown below.
\begin{equation}
\begin{aligned}
x_{n^v,n^p} = &
\left\{
             \begin{array}{ll}
             1, & n^v\,\,is\,\,embedded\,\,on\,\,n^p, \\
             0, & others.
             \end{array}
\right.
\end{aligned}
\end{equation}

The binary variable $y_{N_m^P,N_n^P}^{N_j^V,N_k^V}$ is also used to indicate whether the virtual link $(N_j^V,N_k^V)$ is embedded on the physical link $(N_m^P,N_n^P)$, as shown below.
\begin{equation}
\begin{aligned}
y_{N_m^P,N_n^P}^{N_j^V,N_k^V} = &
\left\{
             \begin{array}{ll}
             1, & (N_j^V,N_k^V)\,\,is\,\,embedded\,\,on\,\,(N_m^P,N_n^P), \\
             0, & others.
             \end{array}
\right.
\end{aligned}
\end{equation}

Each physical node may be embedded by multiple virtual nodes from different VNRs. Expressed as follows.
\begin{equation}
\begin{aligned}
\left\{
             \begin{array}{ll}
             \sum\limits_{n^v \uparrow n^p}x_{n^v,n^p} \geq 1, & n^v \in G_i^V, i=1,2,...,|VNR|, \\
             \sum\limits_{n^v \uparrow n^p}x_{n^v,n^p} = 1, & n^v \in G^V.
             \end{array}
\right.
\end{aligned}
\end{equation}

Each virtual link may be embedded on multiple physical links. Expressed as follows.
\begin{equation}
\begin{aligned}
\sum_{(N_j^V,N_k^V) \uparrow (N_m^P,N_n^P)}y_{N_m^P,N_n^P}^{N_j^V,N_k^V} \geq 1.
\end{aligned}
\end{equation}

If the virtual node $n^v$ is embedded on the physical node $n^p$, the computing resource capacity of $n^p$ should meet the computing resource consumption of $n^v$, which is expressed as follows.
\begin{equation}
\begin{aligned}
CPU_{n^p} \geq CPU_{n^v}, if\,\,n^v \uparrow n^p.
\end{aligned}
\end{equation}

For the physical node $n^p$, the total consumption of computing resource requirements of all virtual nodes embedded in $n^p$ cannot exceed the total computing resource of $n^p$.

\begin{equation}
\begin{aligned}
\sum\limits_{i=1}^{|VNR|}\sum\limits_{n^v \uparrow n^p}CPU_{n_i^v} \leq CPU_{n^p}.
\end{aligned}
\end{equation}

If the virtual link $(N_j^V,N_k^V)$ is embedded on the physical link $(N_m^P,N_n^P)$, the bandwidth resource capacity of $(N_m^P,N_n^P)$ should not be less than the bandwidth resource demand of $(N_j^V,N_k^V)$.

\begin{equation}
\begin{aligned}
BW_{(N_j^V,N_k^V)} \leq BW_{(N_m^P,N_n^P)}, if\,\,(N_j^V,N_k^V) \uparrow (N_m^P,N_n^P).
\end{aligned}
\end{equation}

For the physical link $(N_m^P,N_n^P)$, the total bandwidth resource demand of all virtual links embedded in the physical link $(N_m^P,N_n^P)$ cannot exceed the total bandwidth resource of the physical link $(N_m^P,N_n^P)$.

\begin{equation}
\begin{aligned}
\sum\limits_{i=1}^{|VNR|}\sum\limits_{(N_j^V,N_k^V) \uparrow (N_m^P,N_n^P)}BW_{(N_j^V,N_k^V)_i} \leq BW_{(N_m^P,N_n^P)}.
\end{aligned}
\end{equation}

In SAGIN, the transmission delay of link within different network segments are different. In satellite network, the link delay in the satellite network domain is usually large due to the interference of the propagation medium, radiation and temperature. Moreover, the delay of inter-domain links is often greater than the delay of intra-domain links. We set the delay attribute for physical links and virtual links, and the virtual link can only be embedded on the physical link that is not greater than its delay requirement, as shown below.
\begin{equation}
\begin{aligned}
D_{(N_j^V,N_k^V)} \geq D_{(N_m^P,N_n^P)}, if\,\,(N_j^V,N_k^V) \uparrow (N_m^P,N_n^P).
\end{aligned}
\end{equation}

In the graph model, the transmission of traffic must comply with the law of conservation of traffic, which is a necessary condition for establishing a routing path, i.e., the traffic flowing into the physical node $N_m^P$ must be equal to the traffic flowing out of the physical node $N_n^P$, as shown below.
\begin{equation}
\begin{aligned}
\sum\limits_{N_m^P \in N^P}y_{(N_m^P,N_n^P)}^{(N_j^V,N_k^V)}-\sum\limits_{N_m^P \in N^P}y_{(N_n^P,N_m^P)}^{(N_j^V,N_k^V)} = x_{N_j^V,N_m^P} - x_{N_k^V,N_m^P}, \\
\forall N_n^P \in N^P, \forall (N_j^V,N_k^V) \in E^V.
\end{aligned}
\end{equation}

\subsection{Performance Indicators}

The resource consumption cost of VNE embedded in multi domain SAGIN is calculated as follows.
\begin{equation}
\begin{aligned}
Cost_{G^V \uparrow G^P}=\sum\limits_{i=1,n_i^v \in N^V}^{|n^v|}CPU_{n_i^v}+ \\ \sum\limits_{i=1,(N_j^V,N_k^V)_i \in E^V}^{|e^v|} BW_{(N_j^V,N_k^V)_i} \times hops[(N_j^V,N_k^V)],
\end{aligned}
\end{equation}
where $hops[(N_j^V,N_k^V)]$ represents the number of hops of virtual link $(N_j^V,N_k^V)$.

The goal of VNE is to increase the revenue on the basis of receiving as many VNRs as possible. The revenue of VNE is calculated as follows.
\begin{equation}
\begin{aligned}
Revenue_{G^V \uparrow G^P}=\sum\limits_{i=1,n_i^v \in N^V}^{|n^v|}CPU_{n_i^v}+ \\ \sum\limits_{i=1,(N_j^V,N_k^V)_i \in E^V}^{|e^v|} BW_{(N_j^V,N_k^V)_i}.
\end{aligned}
\end{equation}

We use VNE long-term average revenue, long-term revenue-cost ratio and VNR acceptance rate to evaluate the performance of the SAGIN cross-domain VNE algorithm. The long-term average revenue is calculated as follows.
\begin{equation}
\begin{aligned}
R=\lim_{T \to \infty}\frac{\sum\limits_{t=0}^{T}[Revenue_{G^V \uparrow G^P},t]}{T}.
\end{aligned}
\end{equation}

The long-term revenue-cost ratio is calculated as follows.
\begin{equation}
\begin{aligned}
R/C=\lim_{T \to \infty}\frac{\sum\limits_{t=0}^{T}[Revenue_{G^V \uparrow G^P},t]}{\sum\limits_{t=0}^{T}[Cost_{G^V \uparrow G^P},t]}.
\end{aligned}
\end{equation}

The VNR acceptance rate is calculated as follows.
\begin{equation}
\begin{aligned}
ACC=\lim_{T \to \infty}\frac{\sum\limits_{t=0}^{T}G_{acc}^V}{\sum\limits_{t=0}^{T}G_{arr}^V},
\end{aligned}
\end{equation}
where $G_{arr}^V$ represents the number of virtual network requests arrived, and $G_{acc}^V$ represents the number of virtual network requests successfully embedded.

\section{Algorithm Implementation}\label{part5}

\subsection{Feature Matrix and Policy Network}

The implementation of the SAGIN cross-domain VNE algorithm based on the virtual network architecture is divided into node embedding stage and link embedding stage. We apply the DRL method to the virtual node embedding stage to derive the probability of each SAGIN node being embedded. The key to achieve the desired effect of the DRL method is the effective interaction between the agent and the environment, i.e., the agent needs to be trained in the SAGIN environment as real as possible. Therefore, we use the feature matrix extracted from SAGIN as the input of the agent, and train the agent according to the changes in the underlying resources of SAGIN.

We extract the following four network attributes for each physical node of satellite networks, air networks and ground networks: computing resources, sum of connected link bandwidth, sum of connected link delay and average distance to other non embedded nodes. The above four attributes not only focus on the local characteristics of SAGIN, but also take into account the global characteristics of SAGIN, so they can characterize the underlying network more comprehensively. Among them, the link connected to a physical node refers to the intra domain links, and the sum of bandwidth of the link connected to the physical node is calculated as follows.
\begin{equation}
\begin{aligned}
SUM(n^p)_{BW}=\sum\limits_{(N_m^P,N_n^P) \in E_{n^p}^P} BW[(N_m^P,N_n^P)],
\end{aligned}
\end{equation}
where $E_{n^p}^P$ represents the physical link connected to the physical node $n^p$. In the same way, the sum of delay of the links connected to a physical node is calculated as follows.
\begin{equation}
\begin{aligned}
SUM(n^p)_{D}=\sum\limits_{(N_m^P,N_n^P) \in E_{n^p}^P} D[(N_m^P,N_n^P)].
\end{aligned}
\end{equation}
The larger the value of $SUM(n^p)_{BW}$, it means that when the virtual node $n^v$ is embedded on the physical node $n^p$, a richer bandwidth resource can be selected, and the number of links that can be embedded is more. Conversely, the smaller the value of $SUM(n^p)_{D}$, the smaller the delay interference when the virtual node $n^v$ is embedded on the physical node $n^p$. The average distance to other non embedded nodes in the domain is calculated based on the number of link hops. The smaller the value, the lower the bandwidth resource cost and delay limitation of link embedding. The calculation method is as follows.
\begin{equation}
\begin{aligned}
AVG(n^p)_{DST}=\frac{\sum\limits_{n_i^p \in N^P} DST(n^p,n_i^p)}{|N^P|+1},
\end{aligned}
\end{equation}
among them, $n_i^p$ refers to those physical nodes that have not been embedded by virtual nodes, and $DST(n^p,n_i^p)$ refers to the distance from node $n^p$ to other non embedded nodes in the domain.

It should be noted that the underlying network attributes that can be extracted are far more than the above four. Other network attributes such as node degree, storage resources, etc. are all network attributes that can be extracted. Extracting more network attributes means that more detailed information about the underlying network resources can be provided to the agent, but the computational complexity of the algorithm will also increase. Therefore, after comprehensively considering the actual situation of SAGIN and multi-domain VNE, it is more appropriate to extract the above four features. After extracting the feature of each physical node, the normalized value is concatenated into a feature vector. For nodes in different network segments of SAGIN, the feature vectors are expressed as follows.
\begin{equation}
\begin{aligned}
\left\{
             \begin{array}{ll}
             (CPU(n_s^p),SUM(n_s^p)_{BW},SUM(n_s^p)_D,AVG(n_s^p)_{DST})^T, &\\ n_s^p \in N_S^P, \\
             (CPU(n_a^p),SUM(n_a^p)_{BW},SUM(n_a^p)_D,AVG(n_a^p)_{DST})^T, &\\ n_a^p \in N_A^P, \\
             (CPU(n_g^p),SUM(n_g^p)_{BW},SUM(n_g^p)_D,AVG(n_g^p)_{DST})^T, &\\ n_g^p \in N_G^P.
             \end{array}
\right.
\end{aligned}
\end{equation}

Combine all the feature vectors extracted from SAGIN into a four-dimensional feature matrix. Each row of the matrix is the feature vector of a certain physical node. Agent extracts a feature matrix from SAGIN every time it is trained, so the feature matrix is constantly changing as the underlying network resources change.

\begin{figure}[!h]
\centering
\includegraphics[width=1\columnwidth]{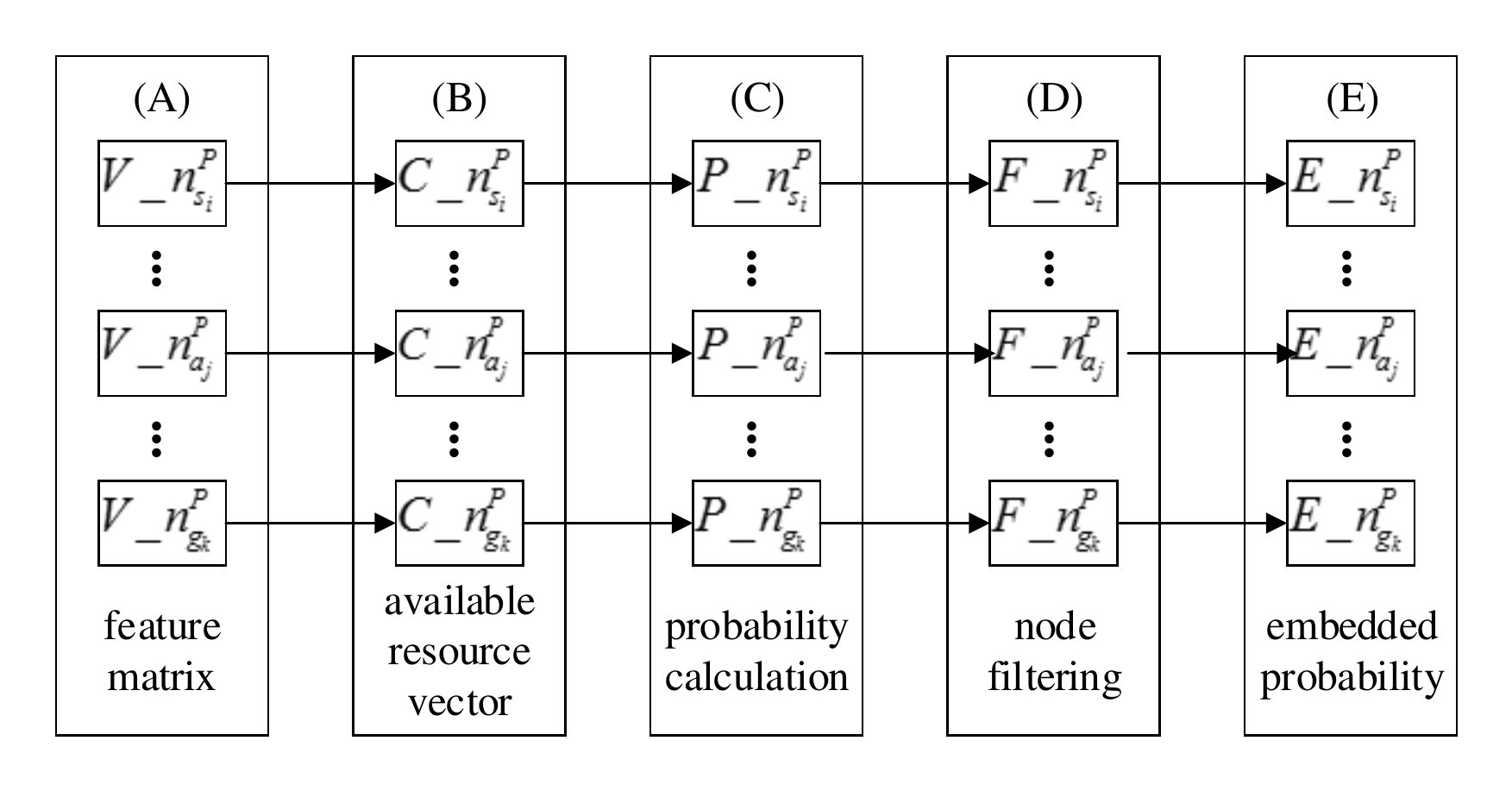}
\caption{Policy network. (A) Extraction layer (B) Convolution layer (C) Probabilistic layer (D) Filtering layer (E) Output layer.}
\label{fig_3}
\end{figure}

The key to DRL method with good perception and decision-making ability is the design and selection of agent. In the proposed algorithm, the agent is assumed by a five-layer policy network, which is composed of the basic elements of neural network. They are extraction layer, convolution layer, probabilistic layer, filtering layer and output layer, respectively. As shown in Fig. \ref{fig_3}. The extraction layer is used to extract the feature matrix from SAGIN during agent training. The convolution layer performs a convolution operation on each feature vector in the feature matrix to obtain the available resource vector form of each feature vector. In probabilistic layer, we use softmax function to generate a probability for each feature vector, i.e., the probability that each physical node is embedded. The filtering layer is used to filter those physical nodes that do not meet the embedding requirements due to excessive resource consumption. The output layer is used to output an available physical node, which is sorted according to the probability of being embedded from large to small. Among them, the convolution operation method is as follows,
\begin{equation}
\begin{aligned}
ARV_i^{cov}=\omega \times v_i + b.
\end{aligned}
\end{equation}

The calculation method of softmax function is as follows,
\begin{equation}
\begin{aligned}
p_i=\frac{e^{ARV_i^{cov}}}{\sum_ne^{ARV_n^{cov}}},
\end{aligned}
\end{equation}
where $ARV_i^{cov}$ represents the $i$-th output of the convolution layer. In this way, the embedded probability of the $i$-th node can be calculated.

\subsection{Training and Testing}

The training of agent is realized in the process of interaction with the environment. Initializing the parameters of policy network, we assume that all VNRs in each VNR period $T$ follow a constant distribution. For each request period $T$, the policy network will extract a feature matrix from SAGIN as input. After the embedding probabilities of all physical nodes are output, the embedding of virtual nodes is completed in a predetermined order. Then we use the breadth first search strategy to complete the embedding of intra-domain links, and finally complete the embedding of inter-domain links.

In multi-domain VNE, we use the revenue-consumption ratio as the agent's reward signal. The revenue-consumption ratio can fully reflect the utilization of the underlying network resources. When the reward signal is large, it means that the node selection strategy currently adopted by the agent can obtain large VNE revenue, i.e., the current action is effective. On the contrary, the agent needs to adjust its actions. The learning rate of agent is also involved in the training phase, and the learning rate will directly affect the gradient of policy network. If the parameter gradient of policy network is large, the training may not derive a better embedding strategy, and the training will be meaningless. In contrast, the training process will be very slow, reducing the efficiency of the algorithm. Therefore, we explore the optimal gradient value by manually adjusting the learning rate. The training process of SAGIN cross-domain VNE algorithm based on virtual network architecture and DRL method is shown in Algorithm 1.

\begin{algorithm}
  \caption{Training}
  \begin{algorithmic}[1]
    \Require
        {$G^P,\,\,G^V,\,\,Policy\,\,network\,\,parameters$};
    \Ensure
        {$Probability\,\,of\,\,SAGIN\,\,nodes\,\,being\,\,embedded$};
    \State {$Random\,initialization\,of\,policy\,network$};
    \While {$iteration<epoch$}
    \For {$n^v \in N^V$}
    \State {$Feature\,\, matrix\,\,extraction$};
    \State {$Probability\,\,derivation$};
    \EndFor
    \If {$isMapped(\forall\,n^v\,\in\,G^V)$}
    \State {$Virtual\,links\,embedding$};
    \EndIf
    \If {$isMapped(\forall\,n^v\,\in\,G^V,\,\forall\,e^v\,\in\,G^V)$}
    \State {$Calculate\,\,reward$};
    \Else
    \State {$Clear\,\,gradient$};
    \EndIf
    \State {$iteration++$};
    \EndWhile
  \end{algorithmic}
\end{algorithm}

In the testing phase, we directly complete the embedding of virtual nodes based on the embedding probability of SAGIN nodes derived from policy network. Then, the breadth first search strategy is used to sequentially complete the embedding of intra-domain links and inter-domain links. The test process is shown in Algorithm 2.

\begin{algorithm}
  \caption{Testing}
  \begin{algorithmic}[1]
     \Require
        {$testset$};
    \Ensure
        {$Three\,\,performance\,\,indexes$};
    \State {$Random\,initialization\,of\,policy\,network$};
    \For {$request \in testset\,\,and\,\,n^v \in request$}
    \State {$Virtual\,\,nodes\,\,embedding$};
    \State {$Using\,\,BFS\,\,strategy\,\,to\,\,find\,\,the\,\,shortest\,\,path$};
    \State {$Virtual\,links\,embedding$};
    \If {$isMapped(\forall\,n^v\,\in\,G^V,\,\forall\,e^v\,\in\,G^V)$}
    \State $return\,(success)$;
    \EndIf
    \EndFor
  \end{algorithmic}
\end{algorithm}

\subsection{Complexity Analysis}

The time complexity of the cross-domain VNE algorithm for SAGIN based on DRL is mainly generated from the two stages of DRL agent training and cross-domain VNE (testing). Since the agent training is performed online and the test phase is performed offline, only the time complexity of the training phase can be considered. For a $4 \times n$ feature matrix, the complexity of extracting it from the underlying network is $O(n)$, and the time complexity of solving all feature vectors is $O(n^2)$. When a new VNR arrives, the feature matrix needs to be updated once, so for all VNRs, the complexity of updating the feature matrix is $O(kn^2)$. Therefore, the time complexity of training stage is $O(n+n^2+kn^2)$, which can be regarded as the final complexity of the algorithm, where $n$ represents the number of nodes in the SAGIN and $k$ is the number of nodes in the successfully embedded VNR.

\section{Experimental Setup and Result Analysis}\label{part6}

\subsection{Simulation Parameters and Environment}

In order to simulate SAGIN, we generate a layered physical network with 100 physical nodes and about 600 physical links, of which 10 physical nodes are used as satellite network nodes, 30 physical nodes are used as air network nodes, and the remaining 60 are used as ground network nodes. There are two inter-domain links connected between each of the three network segments, and the physical nodes connected to the inter-domain links are called boundary nodes. In satellite networks, the computing resources of each physical node are randomly distributed between $20Tflops$ and $40Tflops$, the bandwidth resources of each physical link are randomly distributed between $50Mbps$ and $100Mbps$, and the delay values are randomly distributed between $20ms$ and $40ms$. In air networks, the computing resources of each physical node are randomly distributed between $20Tflops$ and $40Tflops$, the bandwidth resources of each physical link are randomly distributed between $50Mbps$ and $100Mbps$, and the delay values are randomly distributed between $10ms$ and $30ms$. In ground networks, the computing resources of each physical node are randomly distributed between $50Tflops$ and $100Tflops$, the bandwidth resources of each physical link are randomly distributed between $50Mbps$ and $100Mbps$, and the delay values are randomly distributed between $1ms$ and $20ms$. The bandwidth resources of inter-domain links are randomly distributed between $50Mbps$ and $100Mbps$, and the delay values are randomly distributed between $40ms$ and $60ms$.

Besides, we generate 2,000 VNRs, 1,000 of which are used as training set and 1,000 as test set. Each VNR randomly contains 2 to 10 virtual nodes. The computing resource demand of each node is randomly distributed between $1Tflops$ and $20Tflops$, the bandwidth resource demand of each link is randomly distributed between $1Mbps$ and $20Mbps$, and the delay demand is randomly distributed between $1ms$ and $50ms$. The virtual link can only be embedded in the physical link which can meet the bandwidth and delay requirements. Each virtual node will randomly choose which SAGIN segment to embed. We set the batch size to 100, i.e., update the parameters of the policy network once after 100 VNRs, and re-extract a policy network from the underlying network. We summarize the parameter settings of the experimental simulation in TABLE \ref{tab_2}.

\begin{table}
\centering
\caption{Parameter Setting}
\renewcommand\arraystretch{1.5}
\begin{tabular}{|p{50mm}|p{20mm}|}
\hline
Parameter & Value \\
\hline
Physical nodes & 100 \\
Physical links & 600 \\
Satellite network nodes & 10 \\
Air network nodes & 30 \\
Ground network nodes & 60 \\
Computing resources of satellite nodes & U[20,40]$Tflops$ \\
Bandwidth resources of satellite links & U[50,100]$Mbps$ \\
Delay values of satellite links & U[20,40]$ms$ \\
Computing resources of air nodes & U[20,40]$Tflops$ \\
Bandwidth resources of air links & U[50,100]$Mbps$ \\
Delay values of air links & U[10,30]$ms$ \\
Computing resources of ground nodes & U[50,100]$Tflops$ \\
Bandwidth resources of ground links & U[50,100]$Mbps$ \\
Delay values of ground links & U[1,20]$ms$ \\
Bandwidth resources of inter-domain links & U[50,100]$Mbps$ \\
Delay values of inter-domain links & U[40,60]$ms$ \\
\hline
Number of VNRs & 2,000 \\
Number of training sets & 1,000 \\
Number of testing sets & 1,000 \\
Number of virtual nodes & U[2,10] \\
Computing requirements of virtual nodes & U[1,20]$Tflops$ \\
Bandwidth requirements of virtual nodes & U[1,20]$Mbps$ \\
Delay requirements of virtual nodes & U[1,50]$ms$ \\
\hline
\end{tabular}
\label{tab_2}
\end{table}

\subsection{Results Display and Analysis}

\begin{figure*}[!h]
\centering
\includegraphics[width=1\textwidth]{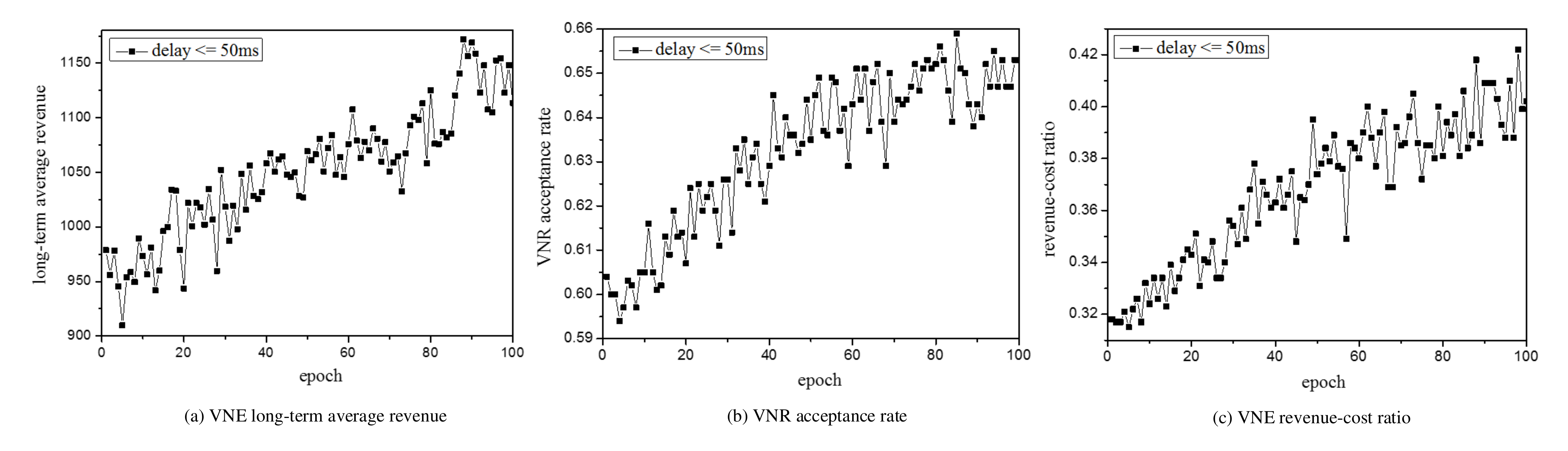}
\caption{The results of the algorithm on the training set.}
\label{fig_4}
\end{figure*}

Since multi-domain VNE is a decision-making problem and is NP-hard, it is necessary to verify the training convergence of the agent under different performance indicators. Fig. \ref{fig_4} shows the changes in agent training from three aspects: VNE long-term average revenue, VNR acceptance rate and revenue-cost ratio. In the initial training stage, since the policy network has just been activated and the agent is not familiar with SAGIN resources, the VNE strategy adopted is random, with low and unstable performance of various indicators. As the training progresses, the agent will continue to explore efficient and reasonable VNE strategies, and the accumulated rewards will continue to increase. After that, the agent will continue to take similar actions to accumulate rewards, so the performance of the three indicators training continues to improve and gradually becomes problematic. Therefore, from the training results, the DRL method based on the policy network is effective.

In order to show that the agent can flexibly adjust the strategy when the environment changes, i.e., the performance of the algorithm may be different when SAGIN resource attribute or the demand of VNR changes, we take the delay factor in VNR as a variable to explore the impact on the algorithm when the user's delay demand changes. In the training phase, the delay requirements of all VNRs in the training set are set according to the pre parameters, which is fixed to $50ms$. The virtual network can only be embedded in the physical network which is not higher than the delay value. TABLE \ref{tab_3} shows the average results of the three performance indicators on the training set when the maximum delay requirements of VNR are $50ms$, $40ms$, $30ms$, and $20ms$. It can be seen that with the continuous improvement of delay performance requirements (the delay value is getting lower and lower), the revenue of VNE and the VNR acceptance rate are significantly reduced. This is consistent with the facts, because with the continuous increase of VNR delay requirements, there are fewer and fewer SAGIN links that can meet the delay requirements, so fewer and fewer VNRs can be successfully embedded, and the revenue and acceptance rate will decline. The revenue-cost ratio does not show a downward trend, because this index has nothing to do with the amount of VNR embedded, it only depends on the revenue and cost of SAGIN resource consumption. Therefore, when the VNR acceptance rate decreases, the revenue cost ratio will not decrease.

\begin{table}
\centering
\caption{Average Performance}
\renewcommand\arraystretch{1.5}
\begin{tabular}{|p{10mm}|p{20mm}|p{20mm}|p{20mm}|}
\hline
- & Average Revenue & Acceptance Rate & R/C \\
\hline
$50ms$ & 1064.929 &	0.644 &	0.343 \\
\hline
$40ms$ & 1069.818 &	0.636 &	0.338 \\
\hline
$30ms$ & 1024.469 &	0.61 &	0.347 \\
\hline
$20ms$ & 894.03 & 0.561 & 0.352 \\
\hline
\end{tabular}
\label{tab_3}
\end{table}

\begin{figure}[!h]
\centering
\includegraphics[width=1\columnwidth]{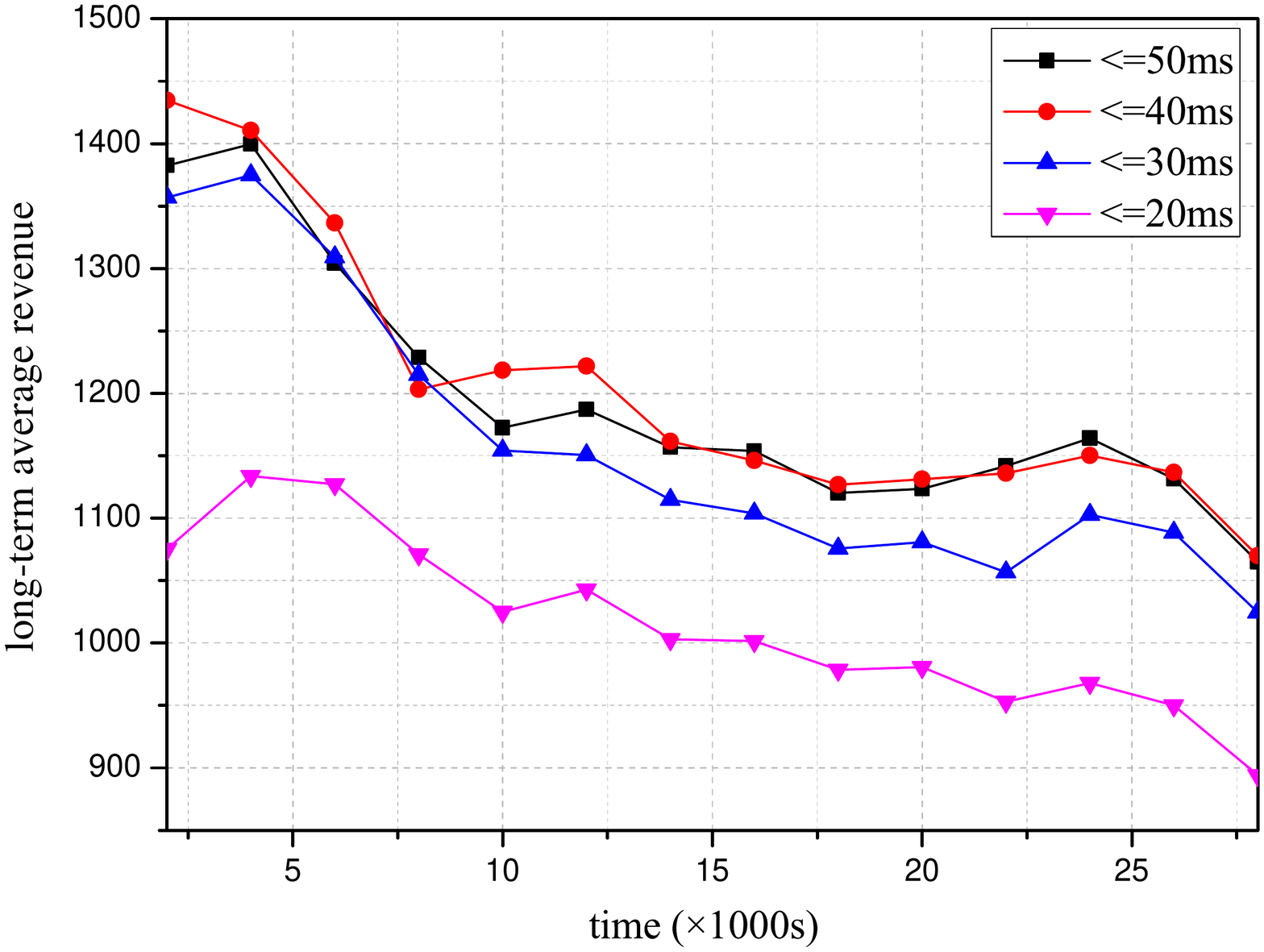}
\caption{VNE long-term average revenue.}
\label{fig_5}
\end{figure}

\begin{figure}[!h]
\centering
\includegraphics[width=1\columnwidth]{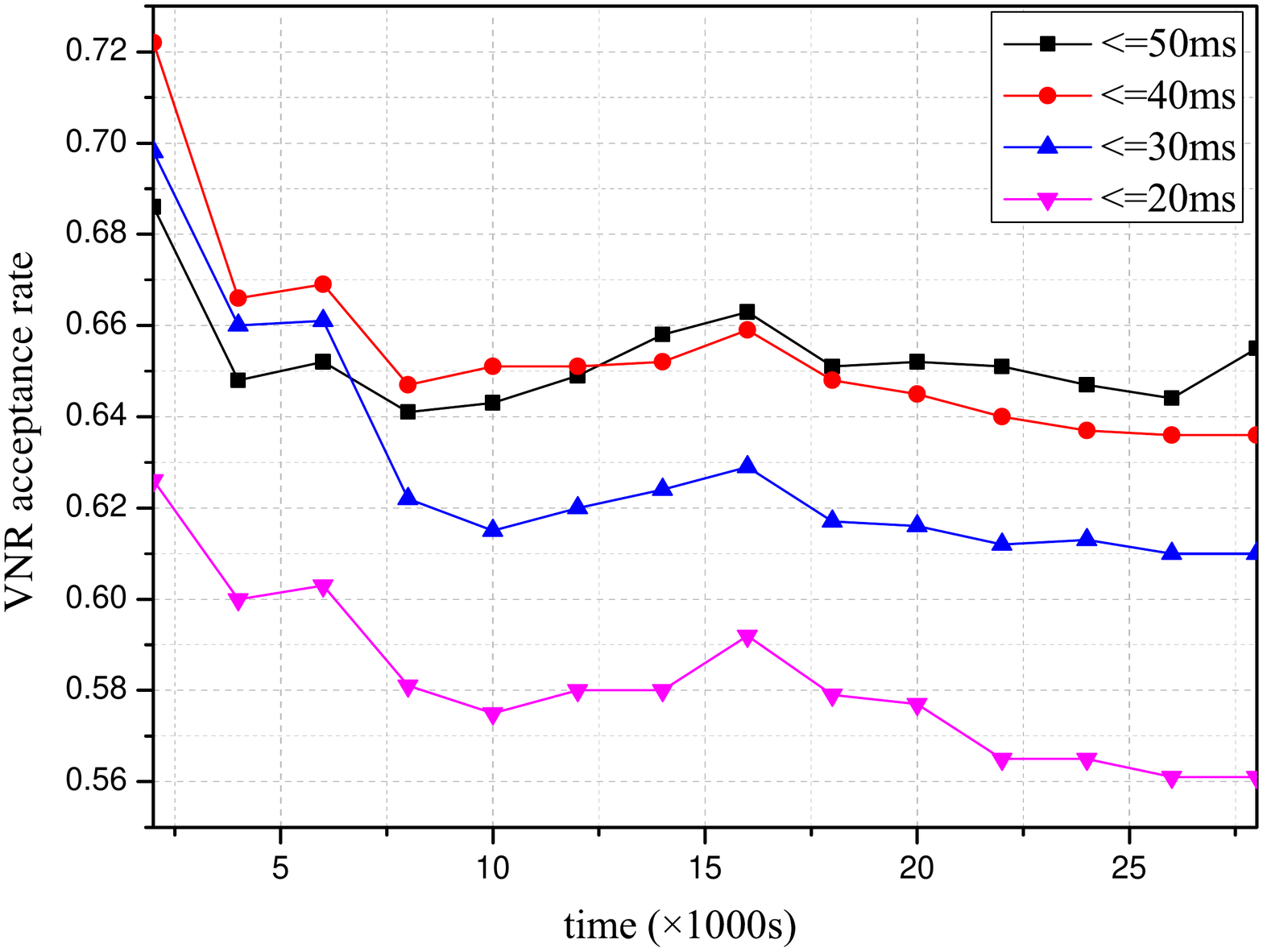}
\caption{VNR acceptance rate.}
\label{fig_6}
\end{figure}

\begin{figure}[!h]
\centering
\includegraphics[width=1\columnwidth]{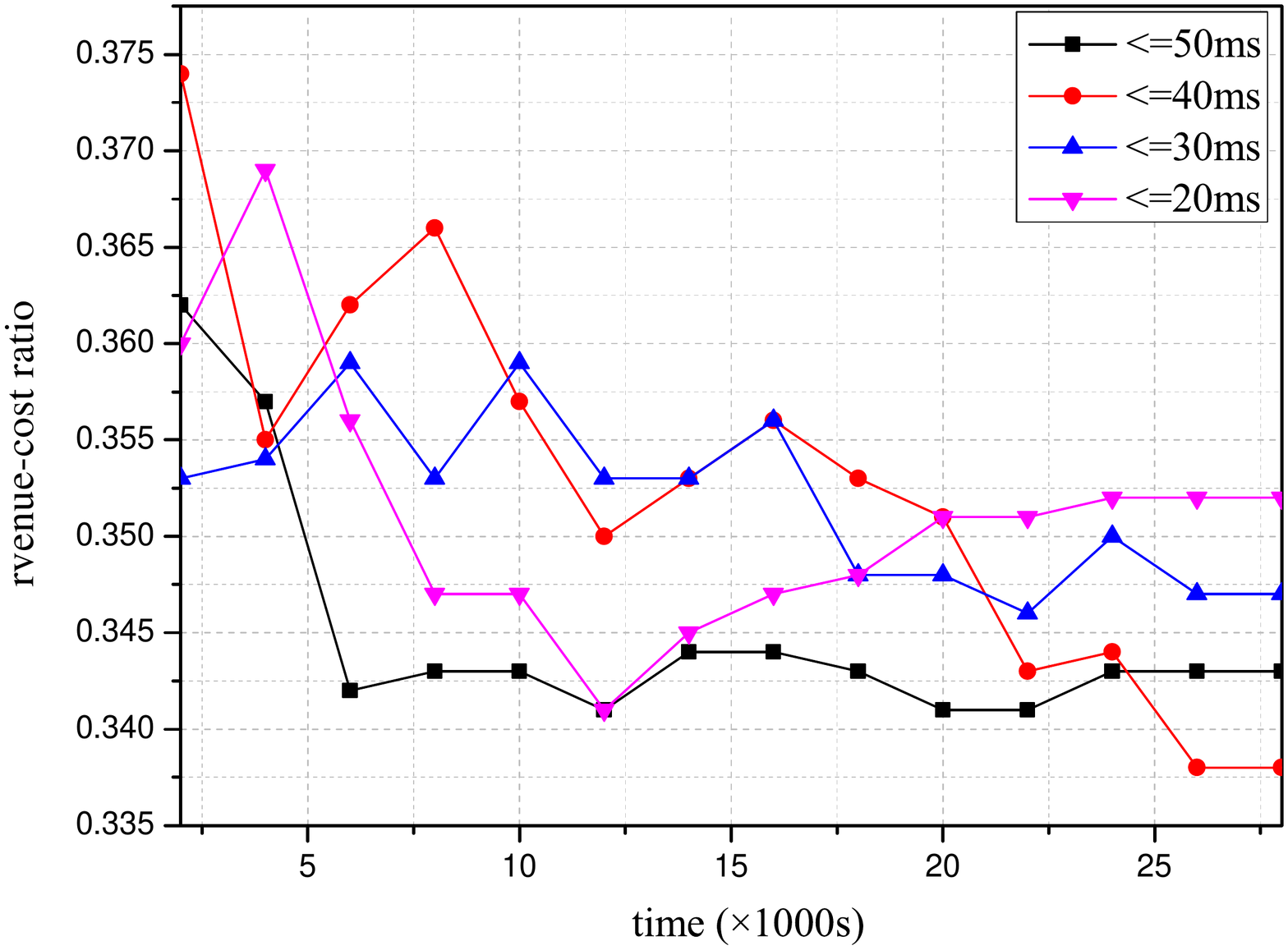}
\caption{VNE long-term revenue-cost ratio.}
\label{fig_7}
\end{figure}

After verifying the effectiveness of the training method, we test the algorithm based on the test set composed of 1,000 VNRs. According to the embedding probability of the physical node derived from training, we directly use the greedy strategy to embed the virtual node \cite{c1}, and then use the breadth first search strategy to embed the virtual link. Fig. \ref{fig_5}, Fig. \ref{fig_6} and Fig. \ref{fig_7} respectively show the test results of the above three indicators under different delay requirements.

From the algorithm test results, it can be seen that the overall changes of the three performances are similar to the training results and are in line with expectations. Since the acceptance rate of VNRs is related to the number of resources of SAGIN, as VNRs continue to be embedded, the available underlying network resources continue to decrease, and the embedding success rate of VNRs continues to decrease. Therefore, the long-term revenue of VNE and the VNR acceptance rate continue to decrease over time. Both VNR acceptance rate and VNE long-term average revenue are affected by changes in delay requirements. As the number of VNRs that can be successfully embedded decreases, both indicators continue to decrease. The test result of VNE revenue-cost ratio shows that this indicator will not change significantly due to changes in delay requirements. In addition, as the delay requirements of VNRs continue to increase, the number of SAGIN nodes and links that can meet the delay requirements decreases, so the performance of the three indicators is continuously reduced. In summary, the experimental results have successfully demonstrated the effectiveness of the DRL-based cross-domain VNE algorithm in the SAGIN resource orchestration field.

\begin{figure}[!h]
\centering
\includegraphics[width=1\columnwidth]{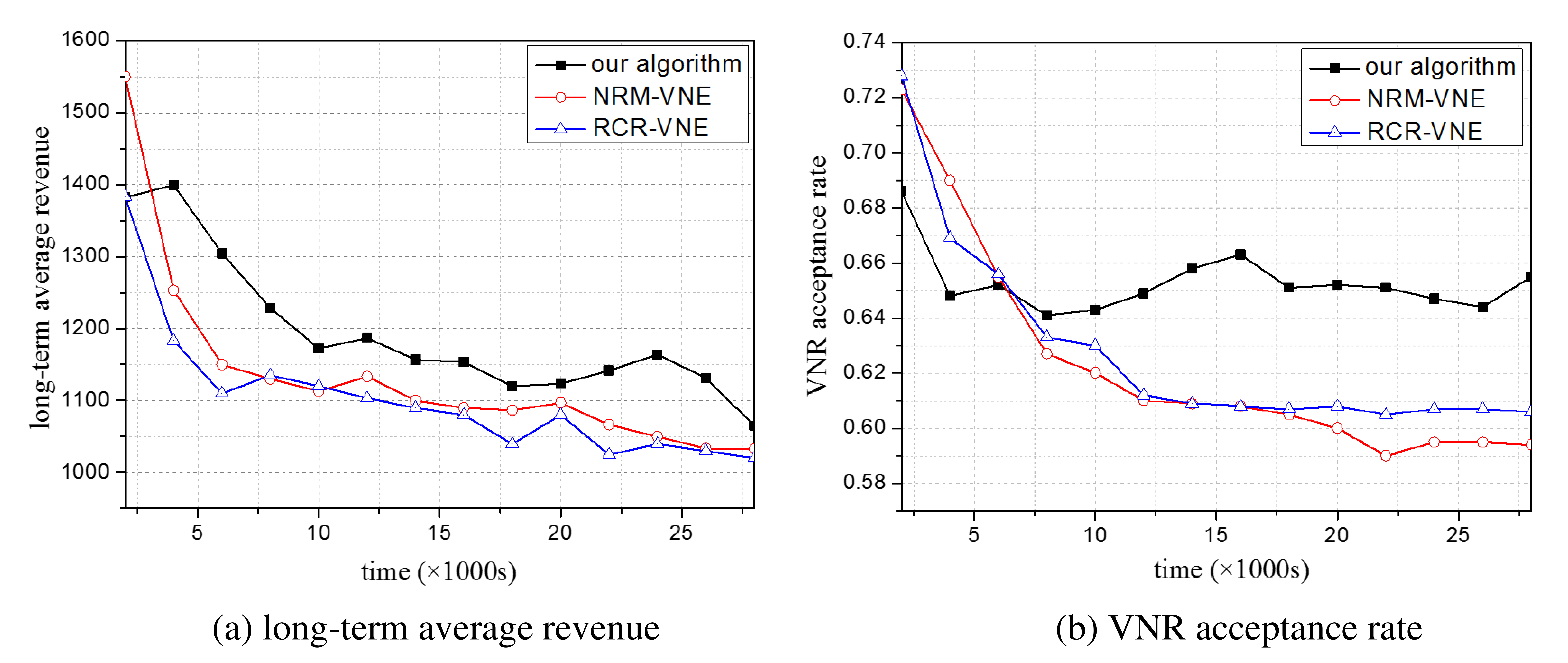}
\caption{Comparison results with benchmark algorithms.}
\label{fig_8}
\end{figure}

In order to further reflect the performance of the algorithm, we compare the algorithm proposed in this paper with the two baseline algorithms proposed in reference \cite{z2}. The SAGIN resource orchestration algorithm based on virtual network architecture is essentially a cross-domain VNE algorithm. In order to ensure the fairness of the comparison, we set the same network resource attributes as this paper for the two comparison algorithms. The MRN-VNE algorithm first calculates the resource metric value of the network node, and then arranges the physical nodes and virtual nodes from large to small according to the value of the value. The virtual nodes complete the mapping in this order. In the link mapping stage, the authors arrange the physical links in order from largest to smallest, and then the virtual link completes the mapping in this order. The RCR-VNE algorithm does not perform the sorting process of virtual nodes. In the link mapping stage, the shortest path algorithm is used to complete the mapping process after sorting according to the bandwidth size. Specifically, we compare the algorithm revenue and virtual request acceptance rate, as shown in Figure \ref{fig_8}.

In general, our algorithm achieves better experimental results than the other two benchmark algorithms. At the beginning of the experiment, because the other two benchmark algorithms adopt greedy strategy, which preferentially selects the physical nodes with abundant free resources for mapping, so their resource revenue and acceptance rate are relatively high. In the subsequent experimental process, the experimental effect of our algorithm is always better than the other two algorithms. On the one hand, NRM-VNE algorithm and RCR-VNE algorithm are two heuristic VNE algorithms. They mainly rely on manual mapping rules (such as setting the sorting method of nodes) to implement VNE algorithm, which greatly limits the flexibility of the algorithm. Our algorithm is a VNE algorithm based on machine learning. This shows that the performance of VNE algorithm based on machine learning method is better than that based on heuristic method. On the other hand, we create a training environment close to the real network for the DRL agent. The agent has fully learned the attributes of SAGIN, so it can make better decisions.

\section{Conclusion}\label{part7}

SAGIN can take advantage of high flexibility, high reliability and high coverage by integrating space network, air network and ground network. However, the seamless integration of the three networks and the orchestration of heterogeneous resources are still a problem. Based on the virtual network architecture, this paper proposes a DRL-based SAGIN multi-domain VNE algorithm. The essence of the algorithm is the allocation of heterogeneous network resources. In DRL, we use the basic elements of neural networks to build a five-layer policy network and use it as the DRL agent. In order to enable the agent to train in real SAGIN environment, we extract four important network attributes for each SAGIN node to form a feature matrix. Through training, the policy network can output the probability of each SAGIN node being embedded. Based on the probability, we complete the VNR embedding in the testing phase. In the experimental phase, we test the performance of the algorithm from both training and testing. In addition, we also analyze the flexibility of the algorithm in dealing with changes in network attributes. Gratifying experimental results show the effectiveness of the DRL-based SAGIN multi-domain VNE algorithm in the arrangement of heterogeneous network resources.

As a part of our future work, we will explore more effective and flexible modeling methods of SAGIN, and set more reasonable resource attributes for network topology. In addition, we will follow the latest research progress in this field, and try to use more comprehensive data to train intelligent agents, so as to obtain better experimental results.

\ifCLASSOPTIONcaptionsoff
  \newpage
\fi

\section*{Biographies}

\begin{IEEEbiography}[{\includegraphics[width=1in,height=1.25in,clip,keepaspectratio]{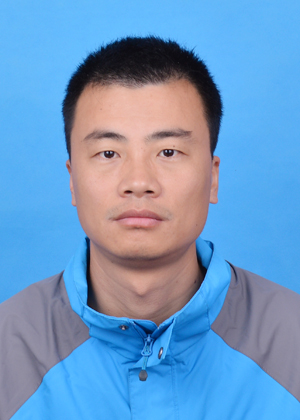}}]{Peiying Zhang}
is currently an Associate Professor with the College of Computer Science and Technology, China University of Petroleum (East China). He received his Ph.D. in the School of Information and Communication Engineering at University of Beijing University of Posts and Telecommunications in 2019. He has published multiple IEEE/ACM Trans./Journal/Magazine papers since 2016, such as IEEE TII, IEEE TVT, IEEE TNSE, IEEE TNSM, IEEE TETC, IEEE Network, IEEE Access, IEEE IoT-J, ACM TALLIP, COMPUT COMMUN, IEEE COMMUN MAG, and etc. He served as the Technical Program Committee of ISCIT 2016, ISCIT 2017, ISCIT 2018, ISCIT 2019, Globecom 2019, COMNETSAT 2020, SoftIoT 2021, IWCMC-Satellite 2019, and IWCMC-Satellite 2020. His research interests include semantic computing, future internet architecture, network virtualization, and artificial intelligence for networking.
\end{IEEEbiography}

\begin{IEEEbiography}[{\includegraphics[width=1in,height=1.25in,clip,keepaspectratio]{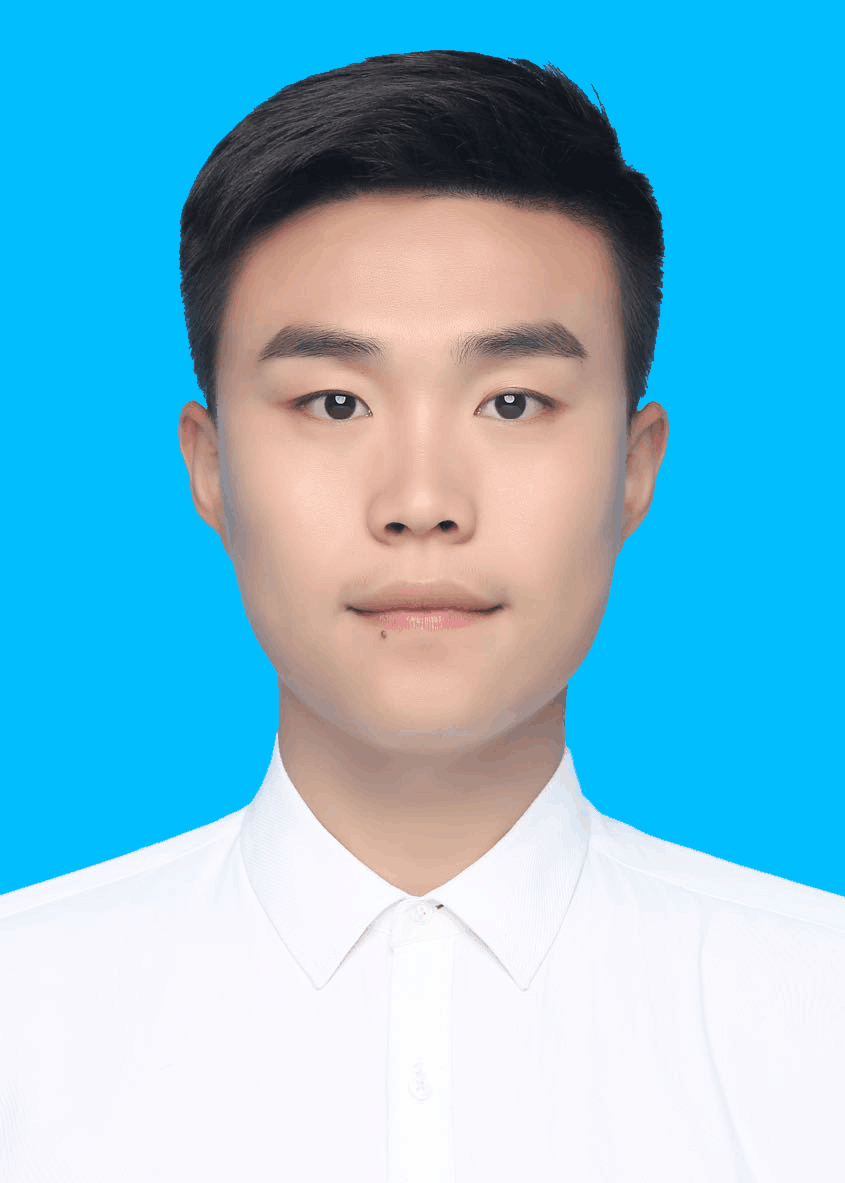}}]{Chao Wang}
is a graduate student in the College of Computer Science and Technology, China University of Petroleum (East China). His research interests include network artificial intelligence, network virtualization and wireless network.
\end{IEEEbiography}

\begin{IEEEbiography}[{\includegraphics[width=1in,height=1.25in,clip,keepaspectratio]{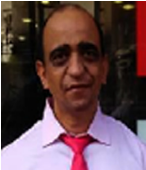}}]{Neeraj Kumar} received his Ph.D. in CSE from Shri Mata Vaishno Devi University, Katra (Jammu and Kashmir), India in 2009, and was a postdoctoral research fellow in Coventry University, Coventry, UK. India. He is working as a Full Professor in the Department of Computer Science and Engineering, Thapar Institute of Engineering and Technology, Patiala, Punjab, India. He is an Adjunct Professor at the King Abdulaziz University, Jeddah, Saudi Arabia and also at the Asia University, Taiwan. He is visiting research fellow at Coventry University, UK. He is a Technical Editor of IEEE Network Magazine, and IEEE Communication Magazine. He is an Associate Editor of IEEE Transactions on Sustainable Computing, ACM Computing Surveys, JNCA, Elsevier, IJCS, Wiley, and Security and Communication, Wiley and on the Editorial Board of Computer Communications, Elsevier. He is senior member of the IEEE. He has more than 8800 citations to his credit with current h-index of 51. He has won the best papers award from IEEE Systems Journal and ICC 2018, Kansas city in 2018. He has edited more than 10 journals special issues of repute and published four books from CRC, Springer, IET UK, and BPB publications. He is Senior Member of the IEEE.
\end{IEEEbiography}

\begin{IEEEbiography}[{\includegraphics[width=1in,height=1.25in,clip,keepaspectratio]{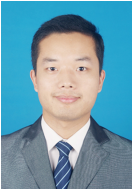}}]{Lei Liu} received the B.Eng. degrees in communication engineering from Zhengzhou University, Zhengzhou, China, in 2010. He received the M.Sc. and Ph.D degrees in communication engineering from Xidian University, Xi’an, China, in 2013 and 2019, respectively. From 2013 to 2015, he worked in a technology company. From 2018 to 2019, he was supported by China Scholarship Council (CSC) to be visiting Ph.D Student in University of Oslo, Norway. He is currently a lecture with the Department of Electrical Engineering and Computer Science in Xidian University. His research interests include, vehicular ad hoc networks, intelligent transportation, mobile edge computing and Internet of Thing.
\end{IEEEbiography}

\end{document}